\documentclass[conference, 10pt]{IEEEtran}
\IEEEoverridecommandlockouts
\usepackage{cite}
\usepackage{amsmath,amssymb,amsfonts}
\usepackage{algorithmic}
\usepackage{graphicx}
\usepackage{textcomp}
\usepackage{algorithm}
\usepackage{multicol}
\usepackage{listings}
\usepackage{color}
\usepackage{skmath}
\usepackage{pgfplots}
\usepackage{setspace}
\usepackage{xcolor}
\usepackage{listings}
\usepackage{textcomp}
\usepackage{xcolor}
\usepackage{multirow}
\usepackage{lipsum}
\usepackage{algorithmic}
\usepackage{here}
\usepackage{multicol}
\usepackage{graphicx}
\usepackage{caption}
\usepackage{subcaption}
\usepackage{setspace}

\definecolor{codegreen}{rgb}{0,0.6,0}
\definecolor{codegray}{rgb}{0.5,0.5,0.5}
\definecolor{codepurple}{rgb}{0.58,0,0.82}
\definecolor{backcolour}{rgb}{0.95,0.95,0.92}
 
\lstdefinestyle{mystyle}{
    backgroundcolor=\color{backcolour},   
    commentstyle=\color{codegreen},
    keywordstyle=\color{magenta},
    numberstyle=\tiny\color{codegray},
    stringstyle=\color{codepurple},
    basicstyle=\footnotesize,
    breakatwhitespace=false,         
    breaklines=true,                 
    captionpos=b,                    
    keepspaces=true,                 
    numbers=left,                    
    numbersep=5pt,                  
    showspaces=false,                
    showstringspaces=false,
    showtabs=false,                  
    tabsize=2
}
 
\def\BibTeX{{\rm B\kern-.05em{\sc i\kern-.025em b}\kern-.08em
    T\kern-.1667em\lower.7ex\hbox{E}\kern-.125emX}}
\begin{document}

\title{Efficient FPGA Implementation of Conjugate Gradient Methods for Laplacian System using HLS}

 \author{\IEEEauthorblockN{Sahithi Rampalli\IEEEauthorrefmark{1},
 Natasha Sehgal\IEEEauthorrefmark{3}, Ishita Bindlish\IEEEauthorrefmark{2}, 
  Tanya Tyagi\IEEEauthorrefmark{4} and Pawan Kumar\IEEEauthorrefmark{5}}
 \IEEEauthorblockA{Center for Security, Theory, and Algorithms Research,
 International Institute of Information Technology, Hyderabad\\
 \IEEEauthorrefmark{1}sahithi.rv1@gmail.com,
 \IEEEauthorrefmark{2}ibindlish@gmail.com,
 \IEEEauthorrefmark{3}natashasehgal289@gmail.com,
 \IEEEauthorrefmark{4}ttyagi96@gmail.com,
 \IEEEauthorrefmark{5}pawan.kumar@iiit.ac.in}}


\maketitle

\begin{abstract}
In this paper, we study FPGA based pipelined and superscalar design of two variants of conjugate gradient methods for solving Laplacian equation on a discrete grid; the first version corresponds to the original conjugate gradient algorithm, and the second version corresponds to a slightly modified version of the same.
 In conjugate gradient method to solve partial differential equations, matrix vector operations are required in each iteration; these operations can be implemented as 5 point stencil operations on the grid without explicitely constructing the matrix. We show that a pipelined and superscalar design using high level synthesis written in C language leads to a significant reduction in latencies for both methods.  When comparing these two, we show that the later has roughly two times lower latency than the former given the same degree of superscalarity.  These reductions in latencies for the newer variant of CG is due to parallel implementations of stencil operation on subdomains of the grid, and dut to overlap of these stencil operations with dot product operations. In a superscalar design, domain needs to be partitioned, and boundary data needs to be copied, which requires padding. In 1D partition, the padding latency increases as the number of partitions increase. For a streaming data flow model, we propose a novel traversal of the grid for 2D domain decomposition that leads to 2 times reduction in latency cost involved with padding compared to 1D partitions. Our implementation is roughly 10 times faster than software implementation for linear system of dimension $10000 \times 10000.$
%
%

\end{abstract}




\section{Introduction and Previous Work}
In this paper, we show an efficient FPGA prototype implementation of conjugate gradient method \cite{saad,demmel,hestenes}. Conjugate gradient method is an efficient iterative technique for solving systems of large, sparse linear equations of the form \[Ax=b,\] where $A \in \mathbb{R}^{n \times n}$ is symmetric and positive definite, and $b \in \mathbb{R}^n,$ the right hand side, are given, and $x \in \mathbb{R}^n$ is the unknown to be determined. Such problems arise frequently in scientific computing, for example, when numerically solving partial differential equations, and in inexact newton methods for solving optimization problems, for example, in machine learning \cite{bottou2017}.

Field Programmable Gate Arrays (or FPGA in short) are becoming increasingly popular due to certain advantages they offer over the more explored and commonly used  processors. First of all, due to their reprogrammable nature, they can be used to develop hardware optimized algorithms before finally implementing them as application specific integrated circuits (ASICs). Secondly, they incur less energy consumption compared to general purpose processors. This the reason why many compute intensive tasks are being implemented as ASIC, an example being the tensor processing unit (TPU) by Google for data analytics (more specifically, to accelarate the inference phase of neural networks \cite{jouppi}).   
Other advantages of FPGAs include explicit fine level control and negligible latency on data movement. In contrast, in shared memory architectures, there are significant latency costs (1000s of cycles) involved in transferring data from RAM to registers. Similarly, in distributed memory architectures, latency costs of inter process communications have significant impact on the scalability of applications. 

Several FPGA implementations of CG have been proposed in the past. Given that the convergence of conjugate gradient method is sensitive to precision used, in \cite{Maslennikow2006}, a fractional number system is proposed. In \cite{Wu2013}, a high performance architecture for CG, in which data blocking to partition large sparse matrices into square blocks, was proposed. This leads to a pipelined and high-throughput sparse matrix vector operation (SpMV) architecture without the need of zero padding for each row in matrices. The authors claim that their architecture can achieve a speedup of upto 9 times compared to a sequential implementation. However, their approach is suitable when the matrices are explicitely available and are stored in off-chip memory. In contrast, our implementation is for problems where matrix vector multiplication is equivalent to a stencil operation. Since, SpMV is a major operation in CG, various implementatons of this operation on FPGAs have been proposed in the  past. For example, in \cite{Zhuo2005}, a binary adder tree is implemented, whose outputs are fed into reduction tree that accumulates partial dot products. However, their design needs zero padding that tends to limit the performance. In \cite{antonio}, Roldao and Constantinides proposed an FPGA implementation of CG suitable for solving series of small to medium size problems. But their design is for dense matrix systems: matrix vector multiplication with dense matrix can be seen as series of dot products, which can be implement using a dot product module (multiplier accumulator model) is used. They studied the trade-off between precision and iteration count of CG. In \cite{R:10} 
the authors present an implementation of dot-product, using a hybrid floating-point and fixed-point number system, describing their acceleration on FPGAs. This paper shows that it is possible to achieve an overall speed of at least 5 times compared to the multi-cycle sequential designs.
In \cite{R:9}, instead of employing floating point operations, this paper focuses on dynamically scaled fixed-point implementation of a preconditioned conjugate gradient method. This allows to attain a higher degree of parallelization. The proposed architecture continuously supplies data operands to a large number of computing units. The architecture was successfully employed for real-time haptic interaction. 
\cite{R:3} implements a highly pipelined conjugate gradient method on FPGA. This paper uses Altera floating point module, which pipelines the typical stages of floating point operations. 

In the papers described above, VHDL or verilog has been used to program FPGAs. In contrast, we use Xilinx Vivado HLS tool (v2017.2) which allows us to program in C and realize the software design on FPGA. By applying simple directives (analogous to OpenMP \cite{openmp} pragmas on multicore machines) to the code, we can achieve similar performance as often achieved by HDL codes with much effort.
 To save on the memory/resource requirement, unlike in  \cite{Wu2013}, we don't assemble the matrix, rather we apply the stencil. Hence, our approach is suitable for solving those problems where SpMV corresponds to applying stencil on a grid. This is true for most linear systems arising from the numerical solution of PDE either via finite difference or finite element methods on structured grid. To achieve superscalarity, we apply a superscalar design of stencil operation on a domain decomposed grid. As has been found in recent literature, certain variants of CG are more amenable for particular hardware. For example, on distributed memory architecture, it is known that interprocessor communication is a major bottleneck in achieving higher scalability. Unlike SpMV that requires only neighbour to neighbour communication, several dot product operations in CG require all to one or so called global synchronization. In \cite{ghysels}, a variant of CG is proposed that reduces the global synchronization latency of dot product by overlapping the dot products with another dot product and a SpMV. In this paper, we show that compared to vanilla CG method that the papers cited above have implemented, the newer variants of the CG methods as proposed in \cite{ghysels} allow higher degree of superscalarity by preventing pipelining stalls.

To implement a superscalar design, we need to partition the domain. In literature, 1D partitioning is typically used. In an 1D partitioning, as the number of partitions increase, the number of points on the boundary increase proportionally compared to the number of points inside the boundary. This leads to degradation in scalability as the superscalarity factor increases for a given problem size, because more time is spent in padding the boundary data. To sove this problem, we propose a 2D partitioning of the grid. For streaming data, we propose a novel grid traversal strategy so that padding of boundary data for each subgrid happens in parallel. To achieve this, we propose the following idea: we first group the subgrids as quadruples in $2 \times 2$ configuration, and the padding of the streaming data starts from the middle corner. The whole grid is partitioned into such quadrapules.    

\begin{algorithm}[h]

\begin{algorithmic}[1]

\STATE $n = \text{size}(A,1)$ 
\STATE $x = \text{zeros}(n,1)$
\STATE $r_{\text{old}} = b$ ; $p = r_{\text{old}}$ ; $j = 1$
\FOR{$i=1$ until convergence}
\STATE $z = A*p$ 
\STATE ${\alpha} = (r_{\text{old}}, r_{\text{old}})/(z, p)$
\STATE $x = x + \alpha * p$ 
\STATE $r = r_{old} - \alpha * z$
\STATE $\beta = (r, r) \, / \, (r_{\text{old}}, r_{\text{old}})$
\STATE $p = r + \beta* p$;
\STATE $r_{\text{old}} = r$
\ENDFOR
\end{algorithmic}
\caption{\label{algoCG} function [A, x, b] = stdCG(A, b)} 
\end{algorithm} 

The following are the main contribution of our paper:
\begin{enumerate}
\item Implementation of a newer variant of CG that allows more pipelining and superscalarity using streaming data. Test results for fairly large grid sizes of $100 \times 100,$ which corresponds to matrix dimensions of $10000 \times 10000.$ Using streaming data allows solving such a large problem.
\item Proposal of a new technique for 2D domain decomposition of the computational grid on streaming data that demands minimal buffering during copying of boundary data to perform parallel stencil operations: A novel traversal of grid, and padding strategy is proposed that allows parallel padding for 2D decomposition of grid well-suited for streaming data as in this paper. 
\item Illustration of high-level synthesis using the well-known C programming language, with suitable directives for achieving pipelining and superscalarity. 
\end{enumerate}

\section{Implementation of the Methods}
We consider the problem of solving the Laplacian system
\begin{align*}
\Delta u = b, \quad \Omega = [0,1] \times [0,1]
\end{align*}
on a 2D unit square domain with zero Dirichlet boundary condition. To solve this problem numerically, we consider an $n \times n$ discrete grid. The derivatives at each of the grid points $(i,j), 1 \leq i,j \leq n$ are approximated by the well known five point stencil \cite{saad}. Since, we are interested in solving large size problems, direct methods such as LU decomposition are not feasible due to large memory requirements and computational costs. \par

\begin{algorithm}[H]
\begin{algorithmic}[1]
\STATE $r = b - A*x$ 
\STATE $w = A*r$
\STATE ${\gamma\_\text{new} = (r,r)}$
\STATE ${\delta_{\text{new}} = (w,r)}$
\STATE $n = A*w$
\STATE $\beta= 0$, \quad $\alpha = \gamma\_\text{new} / \delta$
\STATE $z = n$; \quad $q = w$; \quad $p = r$
\STATE $x = x + \alpha*p$
\STATE $r = r + \alpha*s$
\STATE $w = w + \alpha*z$
\FOR{$i=2$ to convergence}

\STATE $\gamma\_\text{old} = \gamma\_\text{new}$
\STATE ${\gamma\_\text{new} = (r,r)}$

\STATE ${\delta_{\text{new}} = (w,r)}$
\STATE $n = A*w$

	  \STATE $\beta = \gamma\_\text{new} / \gamma\_\text{old}$
      \STATE $\alpha = 1/ (\delta / \gamma\_\text{new}) - \beta / \alpha))$
	\STATE $z = n + \beta*z$
    \STATE $q = w + \beta*q$
    \STATE $p = r + \beta*p$

\STATE $x = x + \alpha*p$
\STATE $r = r + \alpha*s$
\STATE $w = w + \alpha*z$

\ENDFOR
\end{algorithmic}
\caption{\label{algoNewCG} function [A, x, b] = newCG(A, b)} 
\end{algorithm}

\begin{center}
\begin{figure*}
\centering
\begin{subfigure}[b]{\textwidth}
\centering
\includegraphics[scale=.85]{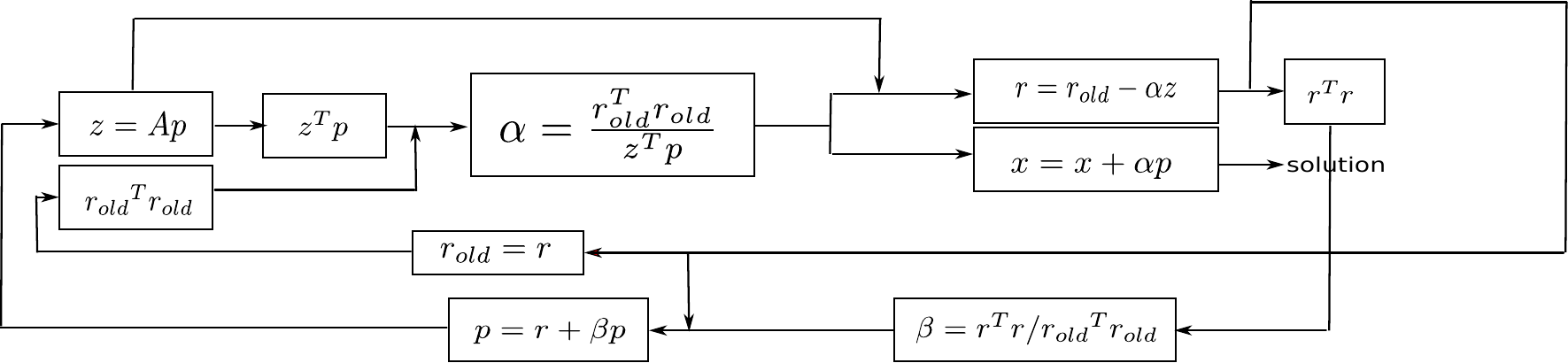}
        \caption{\label{stdCG} Circuit Data Flow Diagram for Conjugate Gradient.} 
\end{subfigure}\vspace{3mm}
\begin{subfigure}[b]{\textwidth}
\centering
 \includegraphics[scale=.85]{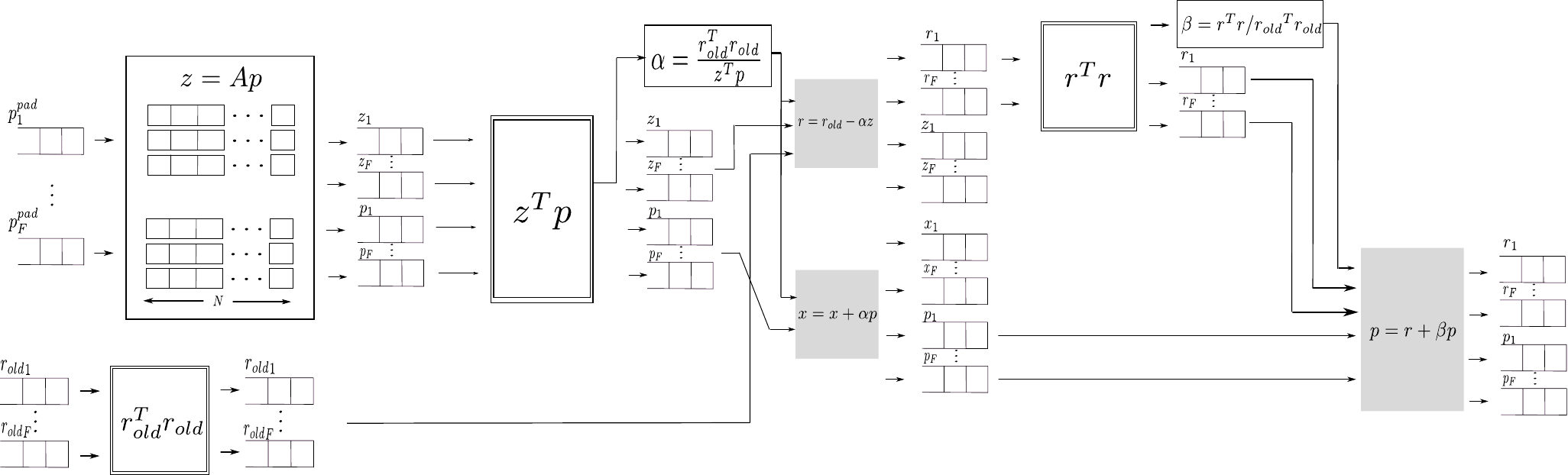}
        \caption{\label{psCG} Block Diagram for Conjugate Gradient.}

\end{subfigure}
\caption{\label{CG} Conjugate Gradient }
\end{figure*}
\end{center}

The conjugate gradient method as shown in Algorithm \eqref{algoCG} is an iterative method of choice for such problems as it requires less memory; about 3 vectors of length $n^2.$ As shown in the Algorithm, the CG method consists of one SpMV in line 5, 3 dot products in line 6 and in line 9, there are 3 axpy operations (vector updates) in lines 7, 8, and 10, and two divisons in lines 6 and 9. In Figure \ref{stdCG}, we show a circuit flow diagram of standard CG. The operations $z=Ap$ and $r_{\text{old}}^T r_{\text{old}}$ can be executed in parallel, and the vector updates $r = r_{\text{old}} - \alpha z$ and $x = x + \alpha p$ can be executed in parallel.

A variant of Conjugate Gradient first proposed in \cite{ghysels} is shown in algorithm \ref{algoNewCG}. We call it NewCG method throughout the paper. It consists of one SpMV in line 15, 2 dot products in lines 13 and 14, 6 axpy operations in lines 18-23, and 4 divisions in lines 16 and 17. In Figure \ref{NewCG}, we show a circuit flow diagram of NewCG. The operations $\gamma_{\text{new}} = (r,r)$, ${\delta_{\text{new}} = (w,r)},$ and $n = A*w$ can be exectued in parallel. The updates of vectors $z, q, p, x, r,$ and $w$ can be executed in parallel. 

Since FPGAs have limited resources, we never construct the full matrix of the Laplacian, because, even when they are written in compact sparse matrix format \cite{saad}, it would require atleast five BRAMs to store the matrix entries. Moreover, since CG solver is used in a larger application, saving BRAM can be used for other necessary data for the problem under consideration.
From the algorithms, we see that conjugate gradient methods essentially have 3 types of operations: SpMV, dot product, and axpy operation. Fig. \ref{psCG} and Fig. \ref{psNewCG} show full pipeline and superscalar structure of CG and NewCG respectively. The single boundary blocks indicate SpMV, double boundary blocks indicate dot products, and shaded blocks indicate axpy operations. The operations overlap in a similar way as shown in Fig. \ref{stdCG} and Fig. \ref{NewCG}. \par

2D Domain Decomposition: In order to parallelize the CG operations and obtain superscalarity, the computational grid is decomposed into blocks, namely quadruple blocks where each quadruple block consists of 2x2 sub-grids. \par
Random access of data elements from a vector is not feasible for a fully pipelined design as load and store operations on storage structures such as BRAMs are limited due to the limited number of ports. Hence, the input to each operation is a stream of data elements of each sub-grid.
In the following subsections, we describe each of the above mentioned operations along with copying of boundary data.

\subsection{ Parallel copying of boundary data}
\begin{center}
	\begin{figure}[H]
	\centering
       \includegraphics[scale=0.85]{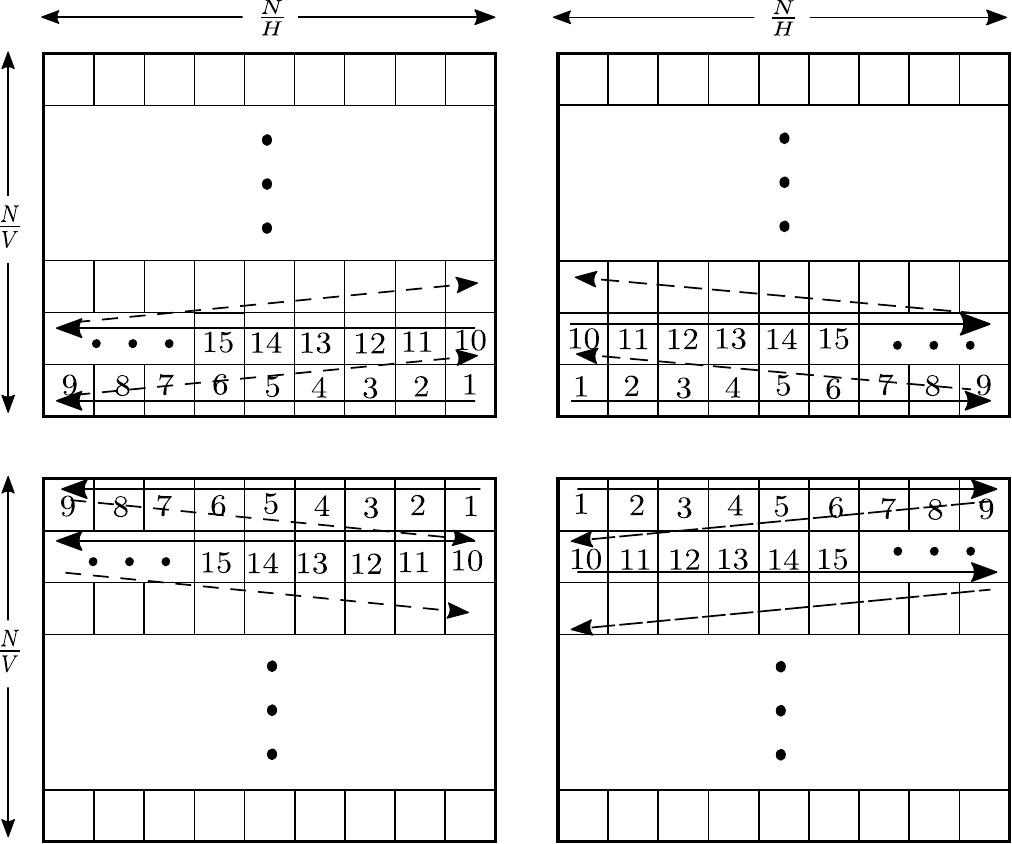}
        \caption{\label{grid1} Streaming pattern in each sub-grid in a quadruple block. Computational grid is divided into $ FACTOR = V \times H$ sub-grids where $V$ is number of vertical partitions and $H$ is number of horizontal partitions. Thus, each sub-grid is of the size $N/V \times N/H$. } 
	\end{figure}
\end{center}

For parallel implementation of stencil operation on each of the sub-grids, we need to copy boundary data from neighboring sub-grids. Since our input data is a streaming data, we propose to traverse the sub-grids from the intersection node of the $2 \times 2$ layout of sub-grids of the quadruple blocks. The traversal pattern is shown in Fig. \ref{grid1}.

\begin{center}
	\begin{figure*}
	\centering
       \includegraphics[scale=0.9]{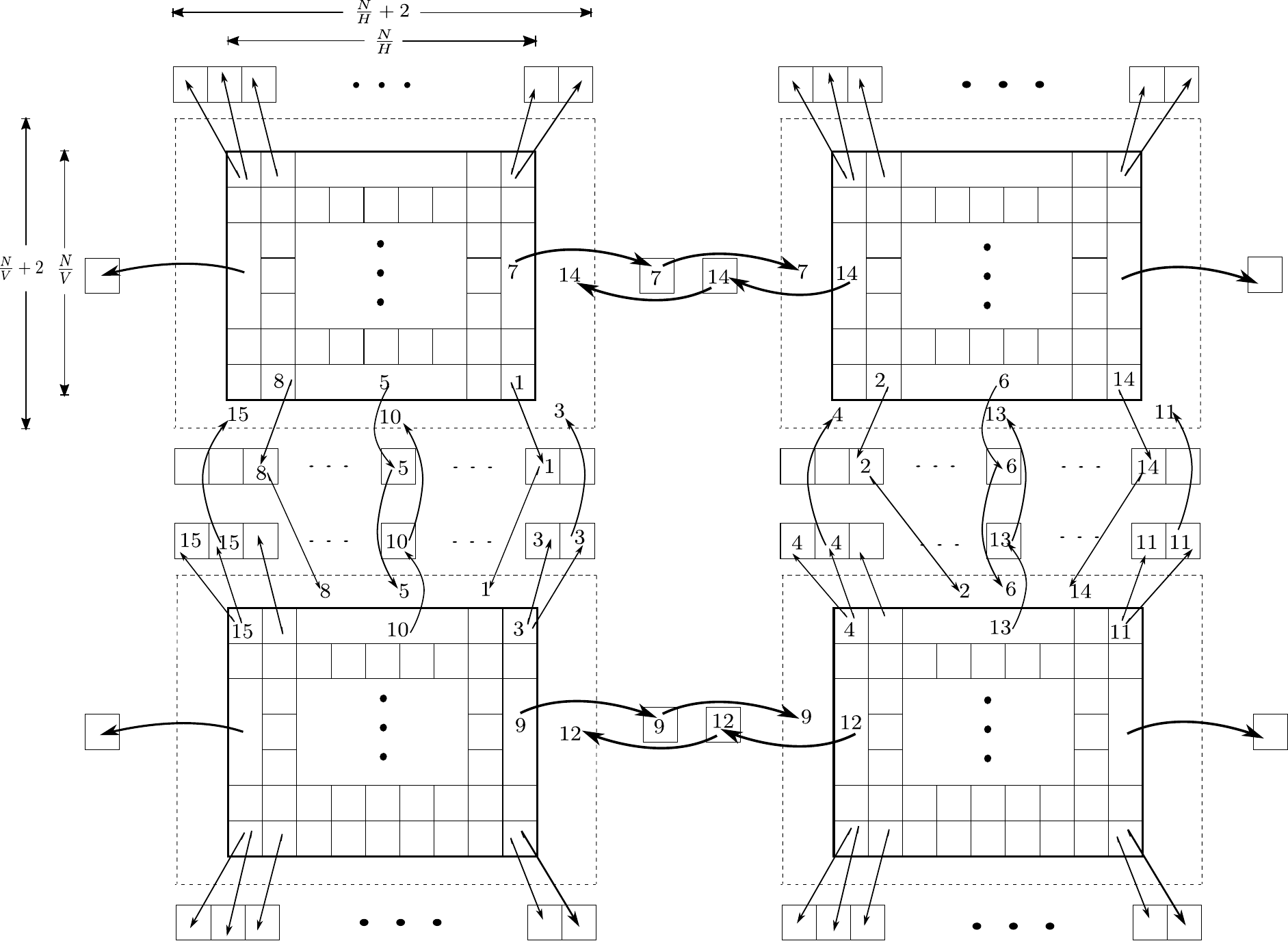}
        \caption{\label{grid2} Copying boundary data across sub-grids in a quadruple block using buffers. Notice that copying not only happens intra-quadruple block but also inter-quadruple blocks. } 
	\end{figure*}
\end{center}

 The traversal in each sub-grid is in the direction of ascending order of numbers. Fig. \ref{grid2} describes copying of boundary data. The copying of numbers is shown for ease of understanding. Since each sub-grid in the quadruple block are of the same size, the exchange of current boundary data happens simultaneously across all sub-grids of all quadruple blocks. This synchronized copying of boundary data requires minimal buffering.

\subsection{Parallel Matrix Vector Multiplication implemented as Stencil operations}
One SpMV operation is required in each iteration of CG. A parallel SpMV operation is achieved by applying local stencil operation on each sub-grid after copying the boundary data. 
 \par
The independent streams of data on which the stencil needs to be applied are cached in line buffers of each sub-grid. 

To perform one stencil operation, multiple data elements of the grid are required. In case of Laplacian system, we have a 5-point stencil which can be conveniently represented as a $3 \times 3$ stencil as follows
\begin{tabular}{|c|c|c|}
\hline
0 & -1 & 0\\ \hline
-1 & 4 & -1 \\ \hline
0 & -1 & 0 \\ \hline
\end{tabular}
. Thus, we require a $3 \times 3$ window of data elements, which are spread across three rows of the grid, to perform a single stencil operation.
A pipelined structure of the stencil operation can be designed with an architecture which consists of line buffers and a window buffer, as designed in \cite{James}. The size of each line buffer is equal to the width of the grid and the size of the window buffer equals that of the stencil. The number of line buffers is equal to the height of the stencil. For example, in our case, we have $3$ line buffers, each of size $n$ and a $3 \times 3$ window buffer. The line buffers use dual port BRAMs, and the window buffer elements are stored in shift registers for parallel access. \par

\begin{center}
\begin{figure*}
\centering
\begin{subfigure}[b]{\textwidth}
\centering
\includegraphics[scale=.85]{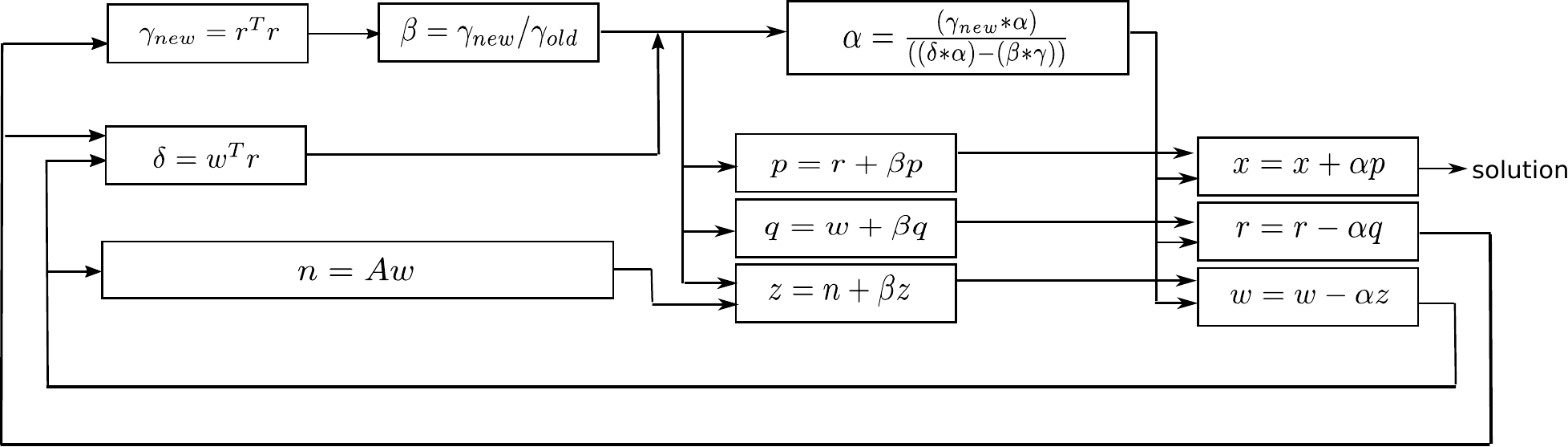}
        \caption{\label{NewCG} Circuit Data Flow diagram for New Conjugate Gradient.} 
\end{subfigure}\vspace{3mm}
\begin{subfigure}[b]{\textwidth}
\centering
       \includegraphics[scale=.90]{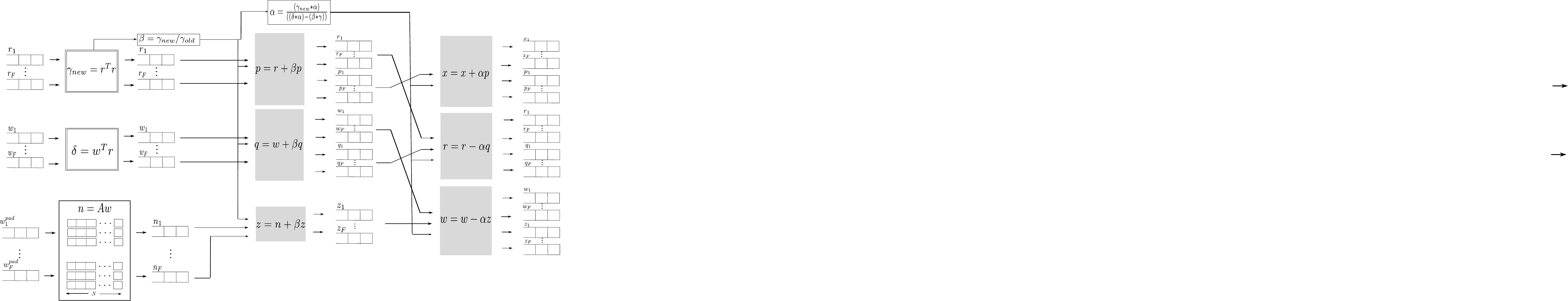}
        \caption{\label{psNewCG} Block diagram for New Conjugate Gradient.}

\end{subfigure}
\caption{\label{newCG2} New Conjugate Gradient }
\end{figure*}
\end{center}

\begin{center}
	\begin{figure}[H]
	\centering
       \includegraphics[scale=0.52]{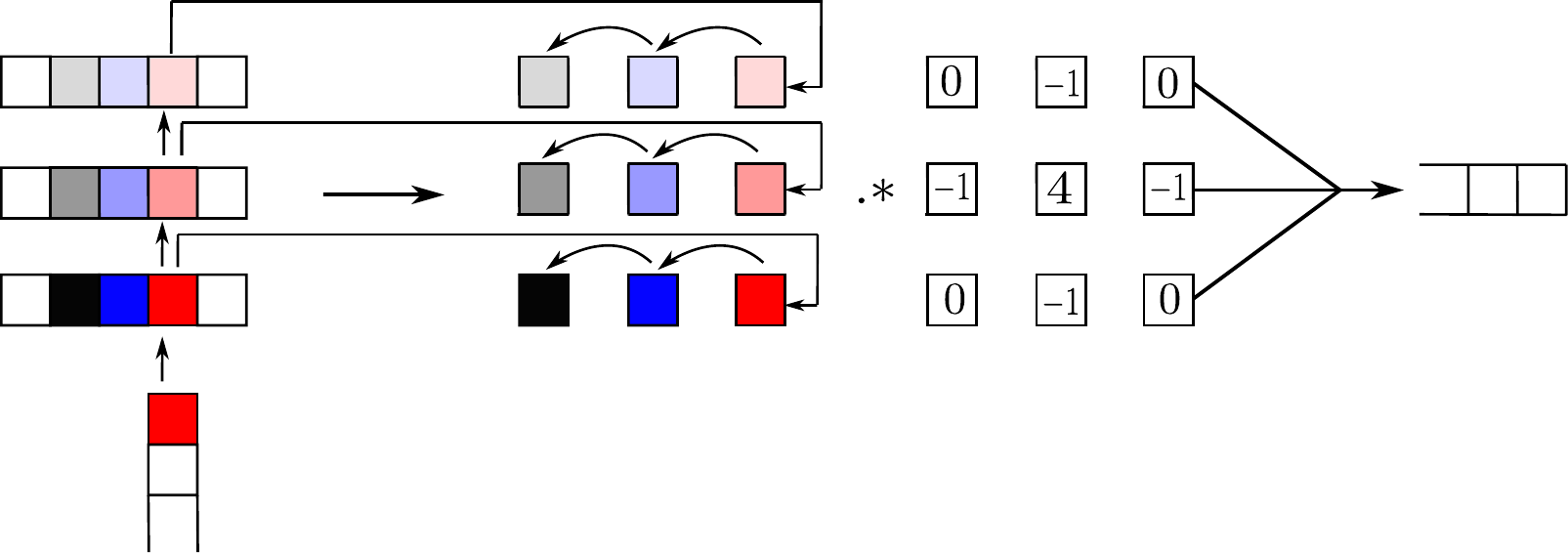}
        \caption{\label{pipeLB} Pipelined Line Buffers} 
	\end{figure}
\end{center}

Figure \eqref{pipeLB} depicts the four stages of the pipeline to perform a stencil operation. The code snippet in Listing 1 shows how a stencil operation is performed using line buffers. 
We use arbitrary precision fixed point type library provided by Xilinx, to represent data elements. We use AP\_RND and AP\_SAT rounding off and saturation modes provided by the library. We specify a fixed point type by $<$T,I$>$, where T and I denote total number of bits and number of integer bits allocated for the element respectively. For example, data elements in listing \ref{ccode} have fixed point precision of $<$50,20$>$ i.e each is allocated 50 bits in total, 20 bits for integral part and 30 bits for fractional parts.
A new data element from the input stream is stored in the last line buffer, as shown in line 18 in Listing 1. When a new element enters a column of the last line buffer, the elements of the column are shifted by one block, upward, as shown in lines 14 to 16. The updated column of the line buffer is copied to the last column of the window buffer in the next clock cycle as shown in line 24. During the same cycle, elements of the window buffer are shifted by one block to the left, as shown in lines 21 to 23. In the next clock cycle, the stencil is applied on the window buffer elements (in Figure \eqref{pipeLB} $.*$ indicates pointwise multiplication of the stencil matrix with the current window buffer elements). The above stages are pipelined as shown in line 11. In the Listing, the variable FACTOR denotes the degree of superscalarity. That is, these pipelined stages are parallelized to FACTOR degrees of superscalarity. \par
The stencil operation begins only when the window buffer is filled with real grid values. 
One stencil operation is done per clock cycle discounting the initial latency. Hence the latency of SpMV for a grid of size $n \times n$ is $(n+2)\times (n/\text{FACTOR}+2)$ $+$ $4$ (The offset 2 is due to the padding of the input stream around the boundary of the local domain).
%
%
\lstset{basicstyle=\small}
\begin{lstlisting}[float=*,language=C, xleftmargin=.07\textwidth, caption= \label{ccode}  HLS implementation of pipelined and superscalar stencil operation.]
typedef ap_fixed<50, 20, AP_RND, AP_SAT> data_t
data_t LineBuffer[FACTOR][3][n / FACTOR + 2];
#pragma HLS ARRAY_PARTITION variable = LineBuffer complete dim = 1
#pragma HLS ARRAY_PARTITION variable = LineBuffer complete dim = 2
data_t WindowBuffer[FACTOR][3][3];
#pragma HLS ARRAY_PARTITION variable = WindowBuffer complete dim = 1
int Stencil[3][3] = {{ 0, -1, 0} {-1, 4, -1} { 0, -1, 0}}
#pragma HLS ARRAY_PARTITION variable = Stencil complete
for(int i = 0; i < (n+2); i++){
	for(int j = 0; j < (n / FACTOR + 2); j++){
	#pragma HLS PIPELINE
		for(int k = 0; k < FACTOR; k++){
		#pragma HLS UNROLL
			for(int p = 0; p<2; p++){
				LineBuffer[k][p][j] = LineBuffer[k][p+1][j];
			}

			LineBuffer[k][2][j] = input_stream[k].read();

			for(int p = 0; p<3; p++){
				for(int q = 0; q<2; q++){
					WindowBuffer[k][p][q] = WindowBuffer[k][p][q+1];
				}
				WindowBuffer[k][p][2] = LineBuffer[k][p][j];
			}

			if( i >= 2 && j >= 2 ){
            	data_t output = apply_stencil(WindowBuffer, Stencil);
            	output_stream.write(output);
            }
		}
	}
}
\end{lstlisting} 

\begin{center}
	\begin{figure}[H]
	\centering
       \includegraphics[scale=.50]{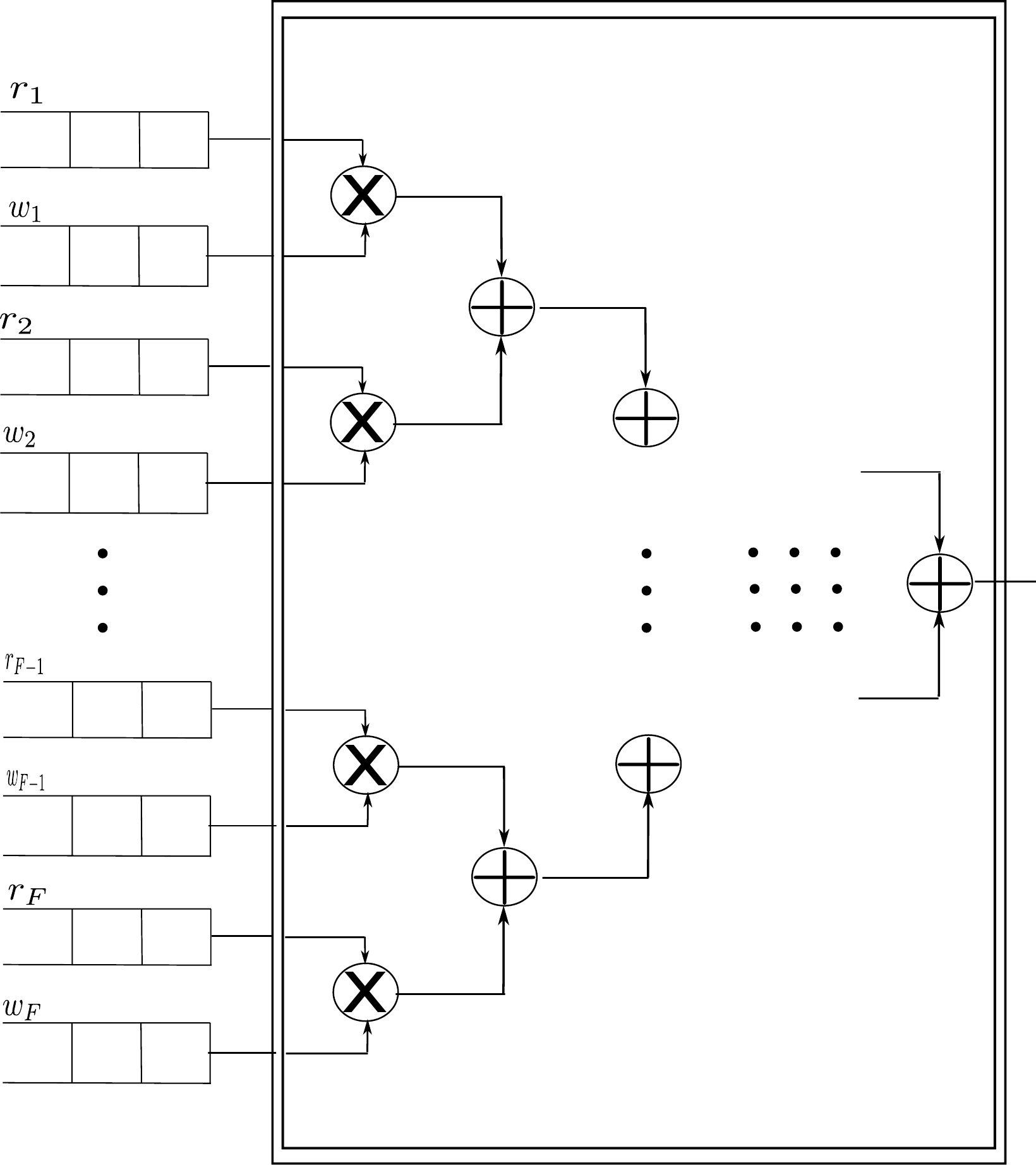}
        \caption{\label{dotprod} Block Diagram for Pipelined and Superscalar Dot Product ($r^Tw$).} 
	\end{figure}
\end{center}

\subsection{Vector Operations}
\subsubsection{Dot Product}
Figure \ref{dotprod} shows how the dot product operation is performed in a pipelined and superscalar design.
FACTOR number of input streams $r_1$ to $r_F$ and $w_1$ to $w_F$ are fed to the dot product operation in parallel. The input streams are processed by a multiplier-adder tree, which has an initial latency of \log[2]{\text{FACTOR}} $+$ $2$.
Let $a$ be the number of cycles for a multiplication operation, and $b$ be the number of cycles for an addition operation, for a given fixed point precision. Then the total latency of the dot product block is \[ a \, \dfrac{n}{ \text{FACTOR}} + b \, \log[2]{\text{FACTOR}} + 2.\]
Let $d$ be the number of DSPs required for a multiplication of given precision. The total number of DSPs used is $\text{FACTOR} \times d$
\subsubsection{AXPY Operation}
\begin{center}
	\begin{figure}[H]
	\centering
       \includegraphics[scale=.70]{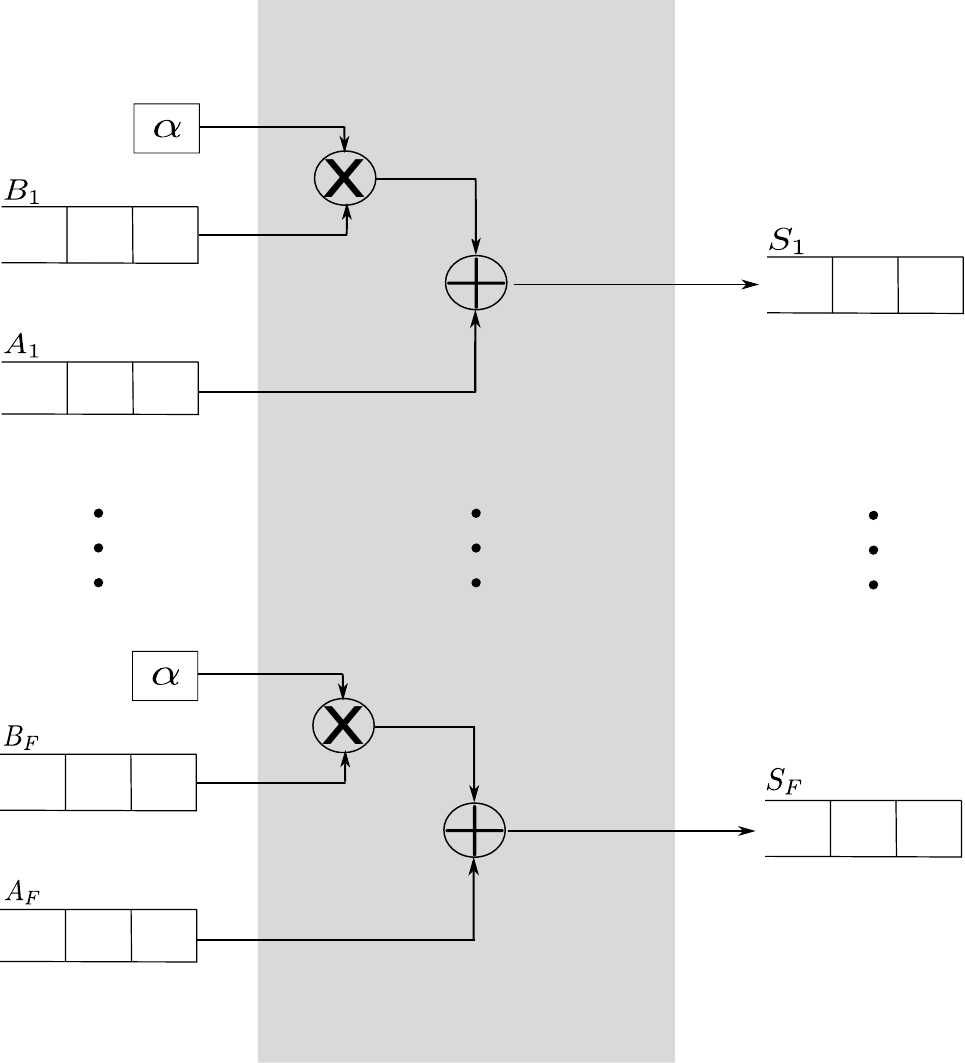}
        \caption{\label{vecADD} Block Diagram for Superscalar axpy Operation($A + \alpha B$).} 
	\end{figure}
\end{center}

Vector-Vector addition is implemented as shown in Figure \ref{vecADD}. Here FACTOR number of input streams $A_1$ to $A_F$ and $B_1$ to $B_F$ are fed to the AXPY operation in parallel. The initial latency is 1 clock cycle. The total latency is approximately given by \[ (a+b) \, \dfrac{n}{\text{FACTOR}}.\]
Number of DSPs used is $\text{FACTOR} \times d.$

\section{Numerical Experiments}
\begin{figure}
\includegraphics[scale=0.65]{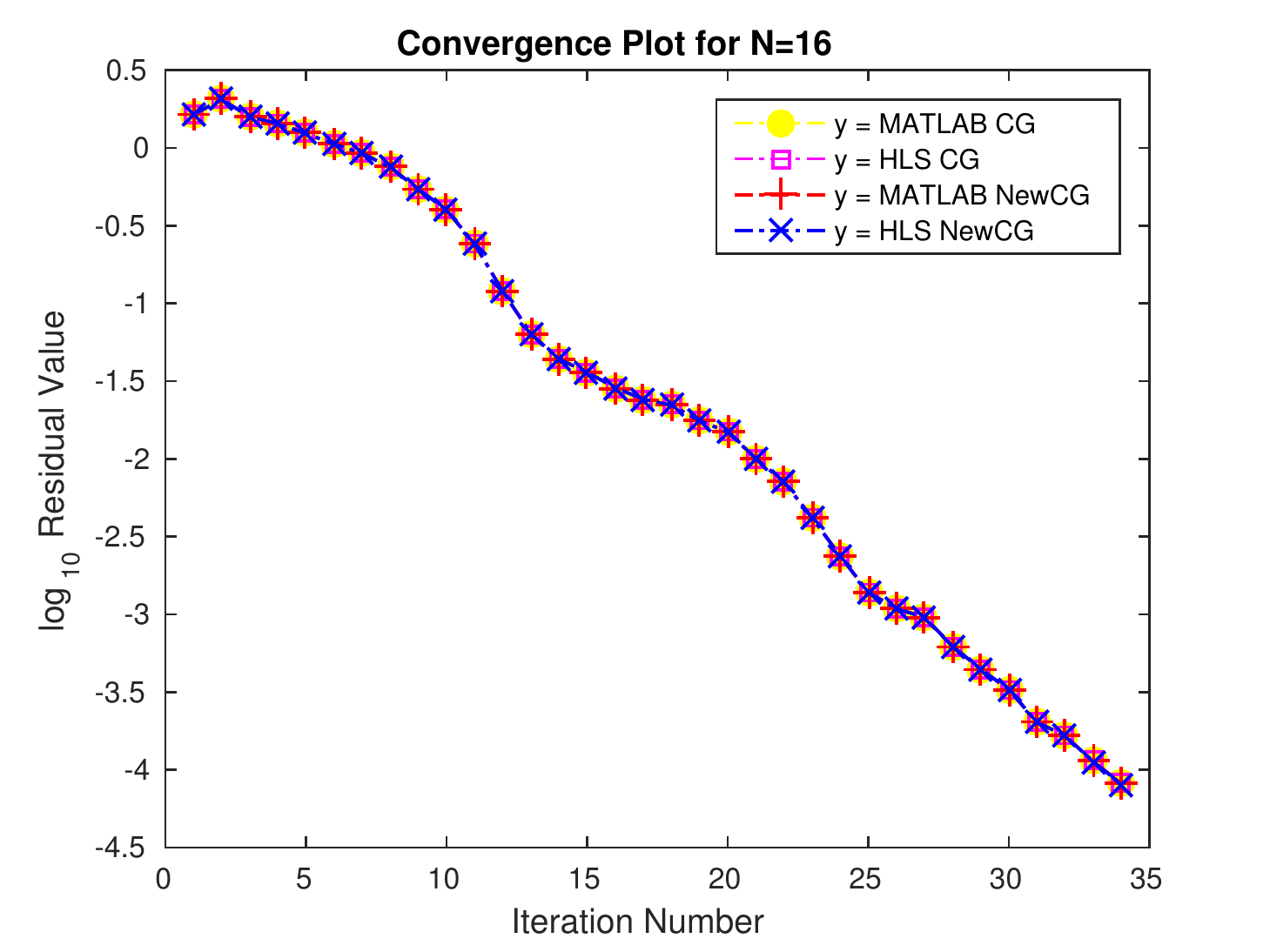}
\caption{\label{convPlots1}Convergence Plot for $N=16$} 
\end{figure}

\begin{figure}
\includegraphics[scale=0.65]{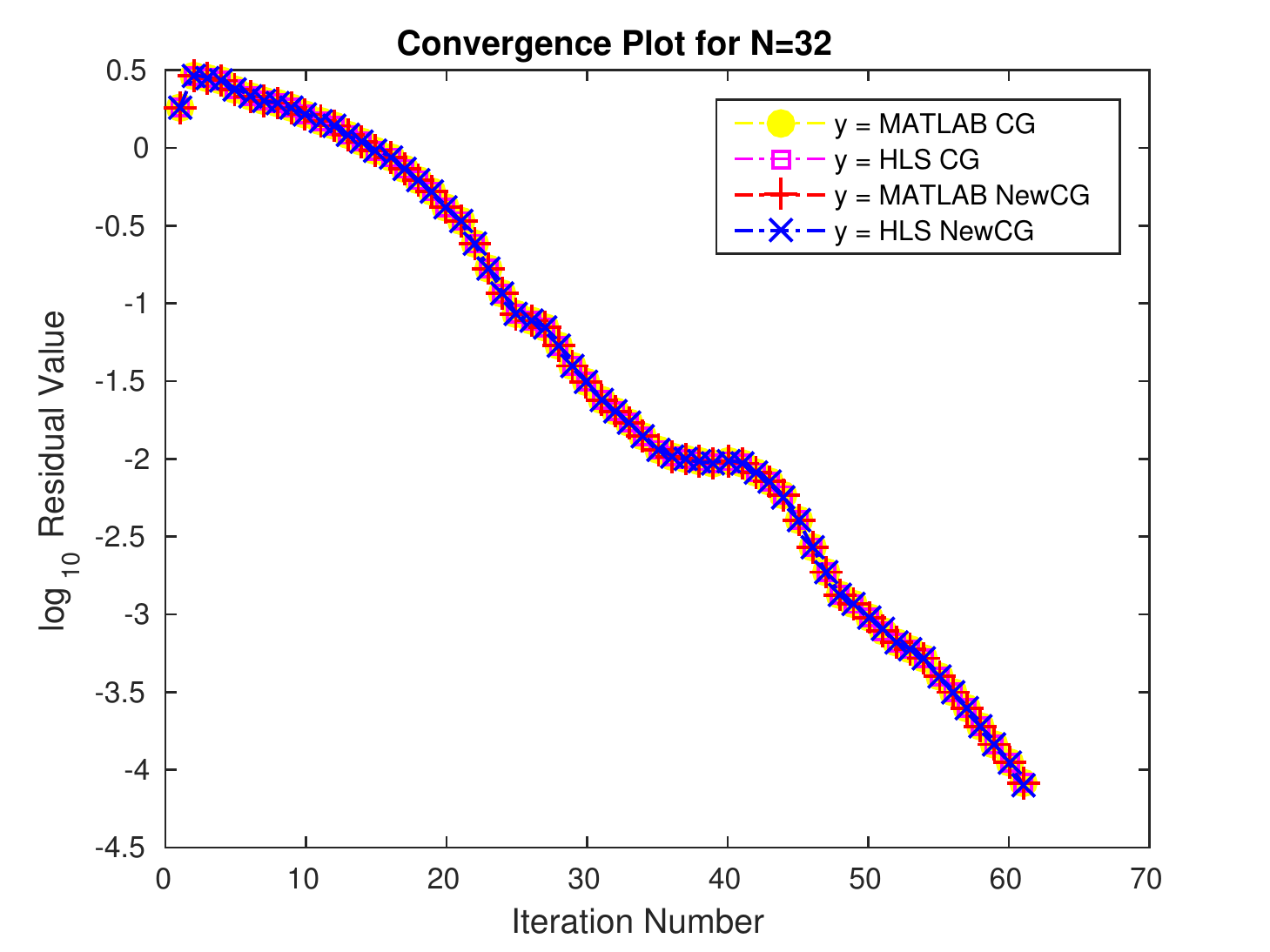}
\caption{\label{convPlots2}Convergence Plot for $N=32.$}
\end{figure}

In Figures \ref{convPlots1} and \ref{convPlots2}, we compare the MATLAB implementations of CG and NewCG with the HLS implementations of CG and NewCG by plotting logarithmic values of relative residuals against the iteration number.

\begin{table}
\begin{center}
 \begin{tabular}{| c  c  c  c |} 
 \hline
$N$ & FACTOR & CG Latency & NewCG Latency \\ 
 \hline
16 &1 & 716 & 360\\ 
16 &  2 & 462 & 232\\ 
16 & 4 & 336 & 179\\ 
32 &  1 & 2253 & 1177\\ 
32 &  2 & 1230& 632\\ 
32 & 4 & 720& 360\\ 
32 & 8 & 466 & 232\\ 
40 &  1 & 3405& 1785\\ 
40 &  2 & 1806& 944\\ 
40 &  4 & 802& 524\\ 
40 &  8 & 610& 314\\ 
40 &  16 & 314& 204\\ 
100 & 8 & 2710& 1424 \\ 
 \hline
\end{tabular}%
\end{center}
\caption{\label{latency} Latency of a Single Iteration of CG and NewCG.} 
\end{table}
We find that the convergence curves of all four implementations are very similar. 
These figures validate the correctnes our HLS implementations of CG algorithms. Note that, while MATLAB uses floating point precision, we use fixed point precision in HLS. The paper \cite{AmbroseXilinx} explains why fixed point type is better than floating point type for obtaining better performance. In particular, floating point addition, being non associative, does not allow superscalarity, as operations are bound to happen sequentially. This issue is eliminated when fixed point type is used. \par

\begin{table}
\begin{center}
 \begin{tabular}{| c  c  c  c |} 
 \hline
$N$ & FACTOR & 1D Pad. Lat. & 2D Pad. Lat. \\ 
 \hline
16 &1 & 562 & -\\ 
16 &  2 & 274 & -\\ 
16 & 4 & 130 & 105\\ 
32 &  1 & 2146 & -\\ 
32 &  2 & 1058 & -\\ 
32 & 4 & 514 & 329\\ 
32 & 8 & 242 & 185\\ 
40 &  1 & 3322 & -\\ 
40 &  2 & 1642 & -\\ 
40 &  4 & 803 & 489\\ 
40 &  8 & 382 & 269\\ 
40 & 16 & 171 & 149 \\ 
100 & 8 & 2350 & 1409\\ 
100 & 16 & 1462 & 734\\ 

 \hline
\end{tabular}%
\end{center}
\caption{\label{padding} Padding latencies for 1D domain decomposition and 2D domain decomposition} 
\end{table}

In Table \ref{latency}, we show how the latencies vary with degree of superscalarity for both the algorithms. With increase in FACTOR, latencies decrease significantly due to an increase in the amount of parallelism at each stage.
We also compare the latency of a single iteration of CG with that of NewCG. A single iteration of CG is approximately twice the latency of NewCG, for all degrees of superscalarity and grid sizes. We find that the larger latency in CG, for a given grid size and FACTOR, is due to the greater number of dot product operations. From  Figure \ref{stdCG}, we can see that the computation of $\beta = (r^T r)/( r_{\text{old}} ^T r_{\text{old}}) $ is stalled by the dot product operation in $r^T r$, which is not overlapped by any other operation that can hide the stalls. 
Also, CG has 2 AXPY operations whereas CHCG has 6 AXPY operations. However, this does not affect the latency of CHCG with respect to CG as 3 AXPY operations of the CHCG are performed in parallel in 2 sets as shown in Fig. \ref{NewCG} \par

In Table \ref{padding}, we compare 1D domain decomposition latencies with 2D domain decomposition latencies during copying of boundary data across sub-grids of all quadruple blocks using the method proposed above. 2D decomposition is applicable only when FACTOR $>$ 2. The padding latencies decrease with increase in FACTOR for a given N. It is evident from the table that latency when 2D decomposition is employed is much less than when 1D decomposition is employed. For larger N, the former performs approximately twice as good as the later performs.\par

Tables \ref{resCG} and \ref{resNewCG} summarize the latencies and the resource utilization for different grid sizes and factors, when all iterations of CG and NewCG are synthesized and 2D domain decomposition is employed. We use square grid dimensions 16, 32, 40, and 100 corresponding to square matrix dimensions 256, 1024, 1600 and 10000 respectively. The trend in latencies for a given N accompanied by an increasing FACTOR, and for an increasing N, is explained by the previously observed trend in single iteration case. This is because the multi-iteration case is a multi-cycle design (finite state automata) of the single iteration case. Thus, the number of DSPs are independent of the number of iterations.
Since each iteration is a pipelined design, the number of DSPs is indepenedent of N. Also, they are approximately proportional to FACTOR, as the number of operators are proportional to FACTOR. This is evident from the dot product and the vector addition block diagrams in Figures \ref{dotprod} and \ref{vecADD}, respectively. The minimal use of BRAMs can be attributed to the streaming data between pipelined stages. \par

\begin{table}[H]
\begin{center}
\addtolength{\tabcolsep}{-3.2pt} 
 \begin{tabular}{|c c c c c c c c c c|} 
 \hline
$N$ & Prec. & $f$ & Iter & Lat & GFLOPs & BR & DSP & FF & LUT \\ 
 \hline
16 & 50,20 & 1 & 33 & 43k &0.33 &0 & 45 & 12k & 10k \\
16 & 50,20 & 4 & 33 & 15k & 0.96&12 & 180 & 41k & 27k \\
32 & 50,20 & 1 & 60 & 27k & 0.39&4 & 45 & 12k & 10k \\
32 & 50,20 & 4 & 60 & 64k &1.63 & 12 & 180 & 41k & 28k \\
32 & 50,20 & 8 & 60 & 40k & 2.62& 24 & 360 & 78k & 51k \\
40 & 50,20 & 1 & 80 & 544k & 0.4 &4 & 45 & 12k & 10k \\
40 & 50,20 & 4 & 80 & 121k& 1.79 &28 & 180 & 41k & 27k \\
40 & 50,20 & 8 & 80 & 71k & 3.05 &24 & 360 & 78k & 50k \\ 
40 & 50,20 & 16 & 80 & 46k & 4.77 &48 & 720 & 153k & 98k \\
100 & 50,20 & 1 & 120 & 4892k & 0.42  &4 & 45 & 13k & 11k \\
100 & 50,20 & 8 & 120 & 499k & 4.09 &56 & 360 & 77k & 51k \\
100 & 50,20 & 16 & 120 & 266k & 7.66 & 112 & 720 & 150k & 98k \\
 \hline
\end{tabular}%
\end{center}
\caption{\label{resCG}Latency and Resource Utilization Results for CG using 2D decomposition. Here Prec. is the fixed point precision $<T,I>,$ $f$ is the FACTOR, Iter is the number of iterations, BR is the number of BRAMs, and Lat stands for latency, i.e., number of clock cycles. One clock cycle is: $10$ nano seconds. Here $k$ denotes order of thousand.} 
\end{table}

From tables \ref{resCG} and \ref{resNewCG}, it is also evident that the latency of NewCG is much less than the latency of CG. However, the decrease in latency is at the cost of greater resource utilization due to more number of parallel operations in NewCG.As number of operations in NewCG are greater than that in CG, DSP utilization is greater in NewCG than in CG. \par
In table \ref{speedup}, we summarize the speedups of maximum FACTOR computed for a given N with respect to FACTOR 1, for both CG and NewCG algorithms. With increasing N, speedup increases for both algorithms. Furthermore, we notice that the speedup of NewCG is better than that of CG for a given N.

For a reference, we also compared with a software implementation of sequential version of CG in SPARSEKIT library \cite{sparsekit}, we find that for $10000 \times 10000$ matrix size and for the same problem, our implementation of NewCG with superscalarity factor of 16 is roughly $15$ times faster.

\begin{table}[H]
\begin{center}
\addtolength{\tabcolsep}{-3.2pt}
 \begin{tabular}{|c c c c c c c c c c|} 
 \hline
$N$ &Prec.&$f$ &Iter&Lat&GFLOPs&BR&DSP&FF&LUT \\ 
 \hline
16 & 50,20 & 1 & 33 & 31k &0.46 &0 & 99 & 19k & 14k \\
16 & 50,20 & 4 & 33 & 10k  &1.47 &12 & 315 & 59k & 38k \\
32 & 50,20 & 1 & 60 & 203k &0.51  &4 & 99 & 20k & 14k \\
32 & 50,20 & 4 & 60 & 42k & 2.45 &12 & 315 & 60k & 38k \\
32 & 50,20 & 8 & 60 & 26k & 4.05 &24 & 603 & 113k & 69k \\
40 & 50,20 & 1 & 80 & 414k  &0.53 &4 & 99 & 20k & 14k \\
40 & 50,20 & 4 & 80 & 81k  & 2.68&28 & 315 & 59k & 38k \\
40 & 50,20 & 8 & 80 & 47k  &4.64 &24 & 603 & 113k & 69k \\ 
40 & 50,20 & 16 & 80 & 29k  & 7.5&48 & 1179 & 219k & 132k \\ 
100 & 50,20 & 1 & 120 & 3718k & 0.59 &4 & 99 & 16k & 20k \\
100 & 50,20 & 8 & 120 & 344k & 5.9 &56 & 603 & 111k & 69k \\
100 & 50,20 & 16 & 120 & 181k & 11.3 &112 & 1179 & 216k & 132k \\ 
 \hline
\end{tabular}%
\end{center}
\caption{\label{resNewCG} Latency and Resource Utilization Results for NewCG using 2D decomposition. Here Prec. is the fixed point precision $<T,I>,$ $f$ is the FACTOR, $Iter$ is the number of iterations, BR is the number of BRAMs, and Lat stands for latency, i.e., number of clock cycles. One clock cycle is: $10$ nano seconds. Here k denotes order of thousand.}
\end{table}

\par

\begin{table}
\begin{center}
 \begin{tabular}{| c | c |c|} 
 \hline
$N$ &  $SP_{CG}$ & $SP_{NewCG}$ \\ 
 \hline
16 & 2.89 & 3.22\\ 
32 & 6.72 & 7.87 \\ 
40 & 11.91 & 14.28 \\ 
100 & 18.38 & 20.59 \\ 
 \hline
\end{tabular}%
\end{center}
\caption{\label{speedup} Speedup of superscalarity of maximum computed FACTOR over that of FACTOR 1, for all grid sizes. $SP_{CG}$ and $SP_{NewCG}$ are speedups in case of CG and NewCG, respectively.  } 
\end{table}

\section{Conclusion}
In this paper, we have implemented a new variant of the conjugate gradient method (CG) that was recently proposed for distributed memory architecture. We have found that the new variant helps in reducing pipeline stalls, compared to the standard conjugate gradient method, because it allows overlapping of dot product and stencil operations. To achieve superscalarity, we implemented 2D domain decomposition. For effective padding of boundary, we proposed a novel grid traversal strategy by organizing subgrids into quadruples, and doing grid traversal in such a way that padding is parallel, and leads to less latency compared to padding for 1D domain domain decomposition that is typically found in the literature.  \par


We have shown that, given a superscalarity factor, new CG has approximately half the aggregate latency cost of standard CG, and roughly $15$ times faster than software implementation of sequential CG.

\bibliographystyle{IEEEtran}
\bibliography{ref}


\end{document}